\documentstyle[preprint,aps]{revtex}

\newcommand{\str}        {$\rm^{1}$}
\newcommand{\gsi}        {$\rm^{2}$}
\newcommand{\legnaro}    {$\rm^{3}$}
\newcommand{\clt}        {$\rm^{4}$}
\newcommand{\bucarest}   {$\rm^{5}$}
\newcommand{\zagreb}     {$\rm^{6}$}
\newcommand{\itep}       {$\rm^{7}$}
\newcommand{\dresde}     {$\rm^{8}$}
\newcommand{\budapest}   {$\rm^{9}$}
\newcommand{\warsaw}     {$\rm^{10}$}
\newcommand{\rrik}       {$\rm^{11}$}
\newcommand{\heidelberg} {$\rm^{12}$}
\newcommand{\seoul}     {$\rm^{13}$}

\begin{document}

\draft

\title{Flow angle from intermediate mass fragment measurements}

\author{
F.~Rami\str,
P.~Crochet\gsi,
R.~Don\`a\legnaro,
B.~de~Schauenburg\str,
P.~Wagner\str, 
J.P.~Alard\clt,
A.~Andronic\bucarest,
Z.~Basrak\zagreb,
N.~Bastid\clt,
I.~Belyaev\itep,
A.~Bendarag\clt,
G.~Berek\budapest,
D.~Best\gsi,
R.~\v{C}aplar\zagreb,
A.~Devismes\gsi,
P.~Dupieux\clt,
M.~D\v{z}elalija\zagreb,
M.~Eskef\heidelberg,
Z.~Fodor\budapest,
A.~Gobbi\gsi,
Y.~Grishkin\itep,
N.~Herrmann\gsi,
K.D.~Hildenbrand\gsi,
B.~Hong\seoul,
J.~Kecskemeti\budapest,
M.~Kirejczyk\warsaw,
M.~Korolija\zagreb,
R.~Kotte\dresde,
A.~Lebedev\itep$^{,}$\rrik,
Y.~Leifels\gsi,
H.~Merlitz\heidelberg,
S.~Mohren\heidelberg,
D.~Moisa\bucarest,
W.~Neubert\dresde,
D.~Pelte\heidelberg,
M.~Petrovici\bucarest,
C.~Pinkenburg\gsi,
C.~Plettner\dresde,
W.~Reisdorf\gsi,
D.~Sch\"ull\gsi,
Z.~Seres\budapest,
B.~Sikora\warsaw,
V.~Simion\bucarest,
K.~Siwek-Wilczy\'nska\warsaw,
G.~Stoicea\bucarest,
M.~Stockmeir\heidelberg,
M.~Vasiliev\rrik,
K.~Wisniewski\gsi,
D.~Wohlfarth\dresde,
I.~Yushmanov\rrik,
A.~Zhilin\itep \\
(FOPI Collaboration) \\ 
}

\address{
\str  Institut de Recherches Subatomiques, IN2P3-CNRS, Universit\'e
Louis Pasteur, Strasbourg, France \\
\gsi Gesellschaft f\"ur Schwerionenforschung, Darmstadt, Germany \\
\legnaro Istituto Nazionale di Fisica Nucleare, Legnaro, Italy \\
\clt Laboratoire de Physique Corpusculaire, IN2P3-CNRS,
Universit\'e Blaise Pascal, Clermont-Ferrand, France\\
\bucarest Institute for Physics and Nuclear Engineering, Bucharest, Romania \\
\zagreb Ru{d\llap{\raise 1.22ex\hbox
{\vrule height 0.09ex width 0.2em}}\rlap{\raise 1.22ex\hbox
{\vrule height 0.09ex width 0.06em}}}er
Bo\v{s}kovi\'{c} Institute, Zagreb, Croatia \\
\itep Institute for Theoretical and Experimental Physics, Moscow, Russia \\
\dresde Forschungszentrum Rossendorf, Dresden, Germany \\
\budapest Research Institute for Particles and Nuclear Physics,
Budapest, Hungary \\
\warsaw Institute of Experimental Physics, Warsaw University, Warsaw, Poland \\
\rrik Russian Research Institute ``Kurchatov", Moscow, Russia \\
\heidelberg Physikalisches Institut der Universit\"at Heidelberg,
Heidelberg, Germany \\
\seoul Korea University, Seoul, Korea\\
}

\maketitle

\begin{abstract}{
Directed sideward flow of light charged particles and intermediate mass
fragments was measured in different symmetric reactions at bombarding
energies from $90$ to $800~{\rm AMeV}$.
The flow parameter is found to increase with
the charge of the detected fragment up to ${\rm Z} = 3-4$ and then
turns into saturation for heavier fragments.
Guided by simple simulations of an anisotropic expanding thermal source,
we show that the value at saturation can provide a good estimate of
the flow angle, $\Theta_{\rm flow}$, in the participant region.
It is found that $\Theta_{\rm flow}$ depends strongly on the impact parameter. 
The excitation function of $\Theta_{\rm flow}$ reveals
striking deviations from the ideal hydrodynamical scaling. 
The data exhibit a steep rise of $\Theta_{\rm flow}$ to a maximum at
around $250-400~{\rm AMeV}$, followed by a moderate decrease as the
bombarding energy increases further.        
}
\end{abstract}

\vspace{0.5cm}


\vspace{1.cm}


\section{Introduction}

The emergence and development of a collective expansion motion
in the course of an energetic heavy-ion collision reflects the response 
of the nuclear system to the internal pressure built-up 
at the interface of the interacting nuclei.
Thus, experimental investigations of this phenomenon are expected
to provide information about 
the compressibility coefficient characterizing 
the nuclear equation-of-state 
which is one of the major objectives in nuclear physics research.
This has generated a widespread interest in this subject which is 
currently the object of intensive experimental activities~\cite{rev 1} 
over a broad range of bombarding energies going 
from the region where a possible liquid-gas phase
transition might occur to the ultra-relativistic energy domain
where the transition to the quark gluon plasma is expected to take place.  

In finite impact parameter collisions, because of the presence of cold 
spectator remnants the expanding nuclear matter in the
hot and dense central region is deflected sidewards along a preferential 
emission direction in the reaction plane.
The most natural experimental observable to characterize this 
effect is the flow angle, $\Theta_{\rm flow}$, {\em i.e.} the angle between the 
direction of the collective sideward motion and the longitudinal beam axis.
This observable is of relevance not only because of its sensitivity to
the stiffness of the nuclear matter equation-of-state, but  
its knowledge is also essential in other measurements such as the 
extraction of the squeeze-out signal around the flow 
axis~\cite{gut 90,cro 96,cro 97}
and two-particle interferometry~\cite{kot 95}.
Its determination requires the use of ``$4\pi$" detectors capable of
global event reconstruction.

The sideward flow can be  characterized 
using the transverse momentum 
analysis method~\cite{dan 85}. The latter method allows one to extract the 
so-called ``flow parameter", $F_{\rm S}$, which is defined as the slope
at mid-rapidity of the average in-plane transverse 
momentum as a function of the rapidity. This observable is 
by definition 
representative of the highly excited and compressed central region. 
It has also the advantage that uncertainties on the reaction
plane determination due to finite number of 
particle effects and detector biases  
can be accounted for by introducing an appropriate correction 
factor~\cite{dan 85,oll 97}. 
An inherent problem in this method is that the flow parameter does not 
reflect only the collective behaviour but it is also sensitive to the
random thermal motion of the emitted particles. 
 In the absence of thermal fluctuations the scaled flow parameter
 ${F_{\rm S}}^{(0)} = F_{\rm S}/{p^{\rm c.m.}_{\rm p}}$, where
 ${p^{\rm c.m.}_{\rm p}}$ is the
 projectile center-of-mass (c.m.) momentum per nucleon,
 is a measure of the flow angle
 (${F_{\rm S}}^{(0)} = \tan(\Theta_{\rm flow})$),
 while in the presence of a large thermal motion the value of
 ${F_{\rm S}}^{(0)}$
 might be significantly lower than $\tan(\Theta_{\rm flow})$.
It follows thereby that
intermediate mass fragments (IMFs), which are less subjected to thermal
fluctuations than do lighter particles, are expected to be strongly 
aligned along the flow direction.
This effect was first observed by the Plastic Ball 
group~\cite{dos 87}
and confirmed later on in several 
experiments~\cite{cro 97,jeo 94,par 95,hua 96,ram 93,ram 97}. 
Recently, quantitative measurements  
of IMF sideward flow in the Kr+Au reaction at $200~{\rm AMeV}$ 
have shown that the flow parameter increases with
the size of the detected particle and reaches a constant limiting value
for fragments with masses $4 \le {\rm A} \le 12$~\cite{hua 96}.
These considerations underscore the importance of accurate measurements
of the flow parameter of IMFs, in particular at beam energies of a few 
hundred ${\rm AMeV}$ where IMFs are copiously 
produced~\cite{ram 93,rei 97,ala 92,her 93}. 

Many interesting processes might simultaneously contribute to the 
observed flow : the release of compressional energy~\cite{sch 74}, 
the thermal pressure~\cite{fri 90}, the momentum dependence of the
nuclear force~\cite{aic 87,gal 87} and the in-medium nucleon-nucleon 
cross sections~\cite{oht 88}. Systematic studies by varying the bombarding 
energy and the system size are needed to disentangle the different   
contributions of these effects since they are 
expected to exhibit different dependences on the initial conditions.  

In the present paper we report on detailed experimental results 
of the flow parameter of light charged particles and IMFs measured 
in different symmetric reactions, Ru+Ru, Xe+CsI and Au+Au, at beam 
energies between $90~{\rm AMeV}$ and $800~{\rm AMeV}$. 
 We show that
 an accurate estimate of the mean flow angle can be obtained
 from the observed flow parameter of heavy IMFs. 
Then, by applying this idea 
we investigate the dependences of the flow angle as a function 
of the collision impact parameter and the bombarding energy.

\section{Experimental setup}

The experimental results reported 
in this paper were obtained in a series of experiments 
carried out using the FOPI 
detector~\cite{gob 93}
at the SIS accelerator facility
of GSI-Darmstadt.
Several symmetric reactions at beam energies going from $90~{\rm AMeV}$
to $1.9~{\rm AGeV}$ were investigated. Here, we will present data for 
the following systems :
Ru on Ru and Xe on CsI at ${\rm E} = 400~{\rm AMeV}$ and Au on Au at 
${\rm E} = 90$, $100$, $120$,
$150$, $250$, $400$, $600$ and $800~{\rm AMeV}$.
The beam intensities were typically ${10}^5$ ions/s. The target thickness
was between $100\;{\rm mg/{cm}^2}$ (at the lowest 
incident energy) and $400 \;{\rm mg/{cm}^2}$ (at the highest energy),
corresponding to an interaction length going from $0.5\%$ to
$2\%$.
Data at $600$ and $800~{\rm AMeV}$ were obtained from the first generation of 
FOPI experiments using the Phase I setup, while the lower beam
energy data were taken more recently with the Phase II configuration
of the FOPI detector.
Details about the FOPI apparatus and its performances 
have been reported in previous publications~\cite{gob 93,rit 95}. 
Here, we recall briefly some of the
sub-detector components of particular interest in the present analysis.

In its Phase I configuration~\cite{gob 93}, the FOPI detector consisted
of a highly segmented Forward Wall of 764 plastic scintillators, 
divided into an external (512 strips) and an internal (252 scintillator
paddles of trapezoidal shape) components. 
To identify the slow  heavy fragments stopped in the external wall
an ensemble of $16$ large gas ionisation chambers with a total of 
$128$ anodes of individual readout, positioned in front of the external
wall, was also used.
For the same purpose the inner part was supplemented with a 
shell of $60$ thin ($2\;{\rm mm}$) plastic scintillator paddles.  
The whole setup covered the laboratory polar angles, $\Theta_{\rm lab}$,
from $1.2^{\circ}$ to $30^{\circ}$ over the full azimuth.
This device allowed us, event by event, to identify the nuclear charge
and measure the vector velocities of most of the light charged
particles and IMFs (up to ${\rm Z}=12$)
emitted in the forward c.m. hemisphere.
In phase II~\cite{rit 95} experiments, 
a cylindrical drift chamber CDC mounted inside a 
superconducting solenoid was used at backward angles 
($\Theta_{\rm lab}=30^{\circ}$ - $150^{\circ}$).
Pions, protons and deuterons were identified in the CDC by means of their
mean energy loss $<{\rm d}{\rm E}/{\rm d}{\rm x}>$ and their laboratory 
momentum, obtained from the 
curvature of the particle tracks in the field of a $0.6~{\rm T}$ strength.
The high granularity of the setup allowed high multiplicity events to be 
measured with a negligible multi-hit rate.
The apparatus ensured also a very good azimuthal symmetry which is an 
important feature for the study of the flow phenomenon.
As the main objective in this work was to extract information on the
sideward flow of IMFs, our analysis was based mainly on the data
from the forward sub-detectors.
The CDC was used only for the purpose of event 
characterization, as
we will see in the next section. 

All data presented in this paper 
were obtained from the analysis of events taken under 
the ``central trigger" condition~\cite{gob 93}.
The latter was defined by adjusting the charged particle multiplicity
to a value which corresponds to impact parameters less than 
$\sim~2/3$ 
of the maximum impact parameter.
At each bombarding energy, samples of about 
${10}^5$ to ${10}^6$   
of such events were recorded.
This large amount of available events allowed us to extract high statistics
data in particular for IMFs.

\section{Event characterisation}

To extract quantitative information on flow phenomena from the
data, one requires a good event characterization both in impact
parameter and azimuth of the reaction plane. 

In order to classify the measured events according to their degree of 
centrality, we have employed the standard method based on the 
correlation between
the multiplicity of the emitted particles and the impact parameter.
The event multiplicity was extracted as the number of charged particles
detected per event in both the outer part (from $7^{\circ}$ to $30^{\circ}$)
of the forward scintillator wall and the CDC.  
For Au+Au at $600$ and $800~{\rm AMeV}$ where the CDC was not operational,
the event multiplicity was restricted only to charged particles
measured by the  outer part of the scintillator wall. 
In the following, the event multiplicity will be labelled MUL when the CDC
is included and PMUL otherwise. 
The measured event multiplicity distributions 
exhibit the typical plateau for intermediate values followed by a rapid
fall off for the highest multiplicities~\cite{rei 97,ala 92}.
These distributions were divided into five intervals,
in accordance to the procedure introduced in previous
works~\cite{ala 92,dos 85}.
The highest multiplicity bin, named MUL5 (or PMUL5) starts at half the 
plateau value and the remaining multiplicity range was divided into four 
equally spaced intervals, named 
MUL1 to MUL4 (or PMUL1 to PMUL4). 
Most of the results which will be presented in this paper deal 
with the MUL4 (or PMUL4) event class corresponding to an average 
geometrical impact parameter of $\sim 3.5~{\rm fm}$ (for the Au+Au reaction).
As we will see, this centrality bin lies in the region where 
the sideward flow reaches its maximum~\cite{ram 97,rei 97}.
The corresponding cross sections and mean geometrical impact parameters
(obtained by assuming a sharp-cut-off approximation) are given in
Tab.~\ref{tab1} and Tab.~\ref{tab2}.
Note in passing that 
the value of the reduced impact parameter is nearly the same 
in all cases which will allow direct comparisons of
the experimental
results obtained for different systems and at different bombarding 
energies.

To reconstruct the reaction plane, we have used the transverse momentum
analysis method~\cite{dan 85}. 
In order to remove autocorrelation effects, the azimuth of the reaction 
plane
was estimated for each particle ${\rm i}$ in a given event
as the plane containing the vector $\vec{Q^{\rm i}}$ and the beam axis 
where $\vec{Q^{\rm i}}$ is calculated
from the transverse momenta $\vec{p_{\rm t}^{\rm j}}$ 
of all detected particles except the particle of interest ${\rm i}$ 

\begin{equation}
\vec{Q^{\rm i}}\;=\;\sum_{{\rm j=1\atop j\neq i}}^{{\rm M}}\,\omega^{{\rm 
j}}
(\vec{p_{\rm t}^{\rm j}} + m^{\rm j} {\vec{v_{\rm b}^{\rm i}}}).
\end{equation}

${\rm M}$ is the multiplicity of the event
and $\omega^{\rm j} = 1\;\mbox{if}\;y^{(0)}>\delta,
-1\;\mbox{if}\;y^{(0)}<-\delta$ and $0$ otherwise.
$y^{(0)}$ is the ${\rm j}^{\rm th}$ particle rapidity divided by the 
projectile rapidity in the c.m. system.
The parameter $\delta$, choosen equal to 0.5, was introduced in order
to remove the contributions of mid-rapidity particles which have a 
negligible correlation with the reaction plane.
According to Ref.~\cite{ogi 89}, a boost velocity
$\vec{v_{\rm b}^{\rm i}} = \vec{p_{\rm t}^{\rm i}}/(m^{\rm sys} -
m^{\rm i}) $
($m^{\rm i}$ is the mass of particle ${\rm i}$ and $m^{\rm sys}$
is the sum of the projectile and target masses)
was applied to each particle ${\rm j}$
in order to take into account the effects of momentum conservation due
to the exclusion of the particle of interest ${\rm i}$.
In order to estimate the accuracy on the reaction plane determination,
due to finite number of particle effects and detector biases,
we have used the method described in reference~\cite{dan 85} which
consists in randomly subdividing each event into two and calculating on 
average the half difference in azimuth, $\Delta {\Phi}_{\rm R}$,
between the reaction planes extracted from the two sub-events.
$\Delta {\Phi}_{\rm R}$ gives an estimate of the
dispersion of the reconstructed reaction plane with respect
to the true one~\cite{dan 85}.
The results are displayed in Tab.~\ref{tab1} and Tab.~\ref{tab2}
in terms of the standard deviation width $\sigma(\Delta {\Phi}_{\rm R})$
extracted from a gaussian fit to the $\Delta\Phi_{\rm R}$ distributions.
As can be seen, the reaction plane is, in all cases, rather well
estimated, with a precision which varies typically from $\simeq 22^{\circ}$
to $\simeq 46^{\circ}$ depending upon the system and the incident energy.
All data subsequently presented in this paper are corrected for these 
uncertainties on the reaction plane determination.

\section{Experimental results and discussion}

In the framework of the transverse momentum method~\cite{dan 85}, the
sideward flow is quantified by plotting the mean in-plane transverse
momentum per nucleon $<p_{\rm x}>$ 
as a function of the rapidity.
For each particle, the quantity $<p_{\rm x}>$ is obtained by 
projecting its transverse momentum
$\vec{p_{\rm t}}$ onto the reaction plane
(the x direction is defined to be in the reaction plane). 
The reaction plane is reconstructed 
by removing autocorrelation effects as described in the previous section.
In the present work, the data have been expressed in terms of 
scale invariant 
(dimensionless)
quantities in order to avoid trivial scaling with the incident 
beam energy~\cite{bon 87}.
In what follows, scaled quantities will be indicated by the
index $^{(0)}$ : 
$<p_{\rm x}^{(0)}> \; = \; {{<p_{\rm x}>} \over {p_p^{cm}}}$ 
will denote the mean in-plane
transverse momentum per nucleon $<p_{\rm x}>$
scaled to the projectile momentum 
per nucleon ${p_p^{cm}}$
in the c.m. system and 
$y^{(0)} \; = \; {{y^{cm}} \over {y_p^{cm}}}$ 
will refer to
the c.m. particle rapidity $y^{cm}$
normalized 
to the c.m. rapidity ${y_p^{cm}}$
of the system. 

Displayed in Fig.~1 are representative examples of 
${<p_{\rm x}^{(0)}>}$ {\em versus} $y^{(0)}$ plots for 
the three reactions Au+Au, 
Xe+CsI and Ru+Ru, all at a beam energy of ${\rm E}=400~{\rm AMeV}$.
The data belong to the multiplicity bin MUL4 where, as outlined earlier,
the directed sideward flow is close to its maximum.      
The ${<p_{\rm x}^{(0)}>}$ values have been divided by the mean cosine
of $\Delta\Phi_{\rm R}$ in order to take into account 
the resolution of the reaction plane determination. 
This has been done according to the method recently proposed by
Ollitrault~\cite{oll 97}. 
The values of $<cos(\Delta\Phi_{\rm R})>$ are given 
in Tab.~\ref{tab1} and Tab.~\ref{tab2}.
The small statistical uncertainties 
(smaller than the symbol sizes for ${\rm Z} = 1$ and  ${\rm Z} = 2$ particles)
on the experimental points, in 
particular for IMFs, illustrate the high statistics recorded in the 
present work. For the Ru+Ru reaction, the data for fragments
with ${\rm Z} > 4$ are subjected to relatively large statistical errors.
There are two reasons for that : i) for this system 
only half of the available statistics was used and ii)   
the production yields of IMFs are lower in lighter systems.
The plots of Fig.~1 were obtained from the analysis of the data taken
by the sub-detector components composing the forward Wall  
(see section II), allowing IMF measurements. This is the reason why
the plots of Fig.~1 are shown only in the forward hemisphere in the
c.m. frame (positive c.m. rapidities).
It is worth recalling that
the forward Wall provides an individual element identification of light 
charged particles and IMFs (up to ${\rm Z}=12$) along 
with velocity measurements.
Thus, the ${<p_{\rm x}^{(0)}>}$ {\em versus} $y^{(0)}$ distribution
measured for a given fragment includes the contributions 
of all associated isotopes. 
It should be, however, stressed that the lack of 
mass identification
does not affect the ${<p_{\rm x}^{(0)}>}$ quantity as it is directly derived
from the measured velocities without any assumption on the mass 
of the particle.

As shown in Fig.~1, the measured ${<p_{\rm x}^{(0)}>}$ {\em versus} $y^{(0)}$ 
distributions exhibit the typical S-shape behavior~\cite{dan 85,dos 86}
reflecting the transfer of momentum between the backward and forward 
hemispheres.
The linear part of the  S-shaped curve is representative of the flow 
in the participant region, the so-called side-splash effect, while 
the fall-off observed at high rapidities is caused by the bounce-off 
effect~\cite{gus 84,gut 89a}. 
It is customary to quantify the magnitude of the participant flow as 
the slope of the $<p_{\rm x}^{(0)}>$ {\em versus} $y^{(0)}$ curve at mid-
rapidity~\cite{dos 86} 
\begin{equation}
F_{\rm S}^{(0)}=\left. {\rm d} <p_{\rm x}^{(0)}> /
{\rm d} y^{(0)}\right|_{y^{(0)} \simeq 0}
\end{equation}
In practice, we extracted this quantity $F_{\rm S}^{(0)}$ 
(known as ``flow parameter" in the literature) by fitting a polynomial
function of the form  
$a + F_{\rm S}^{(0)}\times y^{(0)} +c\times (y^{(0)})^{3}$
to the data.
The fit was restricted to the linear branch of the S-shaped curve.
We have verified that
reasonable changes of the limits of the $y^{(0)}$ range where the fit 
is applied, lead to the same values of $F_{\rm S}^{(0)}$
within the statistical error bars.  
It is worth to notice that the influence of momentum conservation 
effects on the flow parameter was found to be very weak
(less than a few $\%$).
More details on the fitting procedure can be
found in Ref.~\cite{cro 96,cro 97}.

Before 
presenting the results, 
let us briefly comment on the influence of the biases
introduced by our apparatus on the measured values of the flow parameter.
To evaluate these effects, we performed detailed simulations where 
theoretical events were passed through
the FOPI detection filter including geometrical cuts and energy thresholds.
The theoretical events were obtained from Quantum Molecular Dynamics (QMD)
model calculations using the so-called IQMD version~\cite{har 92,har 94}   
which yields generally a quite good agreement with the 
sideward flow data~\cite{wie 93,cro 97b}.
We estimated the effects of detector cuts by comparing the flow parameter
calculated with and without including the experimental filter.
Because of limited statistics (we used 
samples of a few hundred IQMD events per impact parameter unit), 
this could be done accurately only for light particles. 
We found that the $F_{\rm S}^{(0)}$ observable is mainly 
affected by the $\Theta_{\rm lab}=30^{\circ}$ cut. 
The results indicate that acceptance effects depend on the fragment
charge, the collision centrality and the beam energy.  
For MUL4 events at an incident energy of ${\rm E} = 250~{\rm AMeV}$,
$F_{\rm S}^{(0)}$ was found to be biased down by about $30\%$ and $10\%$ 
for ${\rm Z}=1$ and ${\rm Z}=2$ particles, respectively.
For heavier fragments, these effects are expected to be much lower  
as the  $\Theta_{\rm lab}=30^{\circ}$ cut has a weaker influence on IMFs.    
This will be illustrated later on (Fig.~5) using more simple 
simulations based on the decay of an expanding thermal source.

\subsection{Centrality dependence of the flow parameter}

Figure 2 shows the dependence of the scaled flow parameter
measured for Au($400~{\rm AMeV}$)+Au as a
function of the collision impact parameter. 
Error bars correspond to statistical uncertainties multiplied by 
$\sqrt{\chi ^2}$ to take into account the uncertainty from the 
polynomial fit to data.
The impact parameter was obtained from the measured cross sections 
($b_{geo} = \sqrt{\sigma/\pi}$) assuming a sharp-cut-off approximation.
The flow parameter $F_{\rm S}^{(0)}$ is presented here in the form of a 
coalescence invariant quantity~\cite{cro 97b,wan 95,tsa 96}, {\em i.e.} 
including the contributions of all detected particles each weighted by 
its measured charge. 

The results exhibit a maximum of the sideward flow around 
${\rm b} = 4.8~{\rm fm}$.
The observed trend is qualitatively consistent with  the earlier 
results~\cite{dos 86} reported by the Plastic Ball group at the same 
incident energy. 
It can be intuitively understood from the expectation
that the flow parameter should be zero at ${\rm b} = 0$, for symmetry 
reasons, and must tend also towards zero in peripheral collisions.
The value of $F_{\rm S}^{(0)}$ at maximum ($\sim 0.38$) is,
within acceptance effects, in good agreement with the 
data of the Plastic Ball group and those more recently published by 
the EOS Collaboration~\cite{par 95}.

\subsection{Fragment charge dependence of the flow parameter}

The fragment charge dependence of the flow parameter is depicted in Fig.~3
for Au+Au reactions at different bombarding energies. 
Error bars correspond to statistical uncertainties multiplied by 
$\sqrt{\chi ^2}$.
The data are not corrected for the distorsions introduced by
the FOPI apparatus. The influence of detector cuts on the measured 
flow parameters has been already discussed. 
Qualitatively, the data exhibit the same pattern at all five beam energies, 
characterized by a gradual increase of  $F_{\rm S}^{(0)}$ with the 
charge of the detected particle followed by a clear tendency to level-off
above a certain value of ${\rm Z}$. 
With increasing beam energies, the 
satuation appears at lower ${\rm Z}$ values.
This is illustrated by the curves representing the results of a fit with a 
Fermi function to the data.

A similar behaviour has been recently reported in the Kr + Au reaction at 
$200~{\rm AMeV}$
where the flow parameter was found to reach a constant limiting 
value for fragments with masses $4 \le {\rm A} \le 12$~\cite{hua 96}.
It is worth-while mentioning that this trend has been also seen for 
the squeeze-out effect~\cite{cro 96,bas 97}  and seems to be typical 
for all flow observables~\cite{ram 97}.  

The increasing strength of the sideward flow with the fragment size
was observed for the first time~\cite{dos 87}
by the Plastic Ball group for Au($200~{\rm AMeV}$)+Au reactions and confirmed,
since then, in several  
experiments~\cite{cro 97,jeo 94,par 95,hua 96,ram 93,ram 97}. 
This phenomenon was predicted by hydrodynamical models~\cite{sto 81}
where collective effects were found to be much more visible for heavier
mass fragments. QMD calculations also yield larger flow for heavier
fragments~\cite{pei 89}. 
This large flow carried by fragments is of particular interest because
of its enhanced sensitivity to the parametrisation of the nuclear
equation-of-state used in dynamical model 
calculations~\cite{cro 97b,pei 89,zha 95}.  

It can be also understood from the interplay between 
collective and thermal (random) motions. 
In an idealized   picture in which nucleons and fragments are emitted 
from a common expanding thermalized source, heavy fragments are less 
subjected to 
thermal fluctuations than do lighter particles. Thermal fluctuations are
governed by the thermal energy which is independent of the mass of the 
particle, while the collective expansion energy increases linearly with
the mass and is, therefore, better reflected in heavier fragments.
Thus, it is easy to understand, within this simple picture, that in 
the presence of thermal fluctuations the apparent flow angle 
that one can extract from the flow parameter 
$\Theta^{\rm app}_{\rm flow} = \arctan({F_{\rm S}}^{(0)})$ 
is lower than the effective flow angle. 
The nearly constant value observed in the flow parameter 
of heavy fragments (Fig.~3) can be attributed to the fact that above a 
certain mass (or charge) the emitted fragments are only very little affected 
by the random thermal motion so that their apparent flow angles are 
expected to be very close to the effective flow angle. 
That is the idea that we will exploit below (section IV.D) in order
to extract an estimate of the flow angle from the data.

It should be mentioned that it was shown in Ref.~\cite{wan 95}
that the increase of the sideward flow with 
fragment size can be also described by a simple momentum space
coalescence prescription providing a transverse momentum cut
of 0.2~AGeV is imposed. 

\subsection{System size dependence of the flow parameter}

In Fig.~4, the normalised flow parameter $F_{\rm S}^{(0)}$ as a function
of the charge of the detected fragment is displayed for the
three reactions under study. 
The flow parameter is divided here by the quantity 
${\rm A}_{\rm P}^{1/3}+{\rm A}_{\rm T}^{1/3}$, where 
${\rm A}_{\rm P}$ and ${\rm A}_{\rm T}$ 
denote the mass of, respectively, the projectile and target nuclei.
Two observations can be readily made from the examination of this figure. 
First, the dependence of the flow parameter on the fragment charge measured  
in Ru+Ru and Xe+CsI reactions follows the same trend as observed in the
heavier system Au+Au. 
Second, the projectile-target mass dependence
of $F_{\rm S}^{(0)}$ is consistent, within statistical uncertainties,
with the ${\rm A}_{\rm P}^{1/3}+{\rm A}_{\rm T}^{1/3}$ empirical 
scaling rule introduced 
recently by the EOS collaboration~\cite{cha 97}. 
The origin of this scaling rule is not yet well understood. 
It might be attributed, in a hydrodynamical picture, to the
fact that for collisions with velocities well above that of sound, the 
pressure built-up should scale with collision 
length or time~\cite{cha 97}. 
  
\subsection{Extraction of the flow angle from IMF measurements}

As outlined in the introduction, the most natural observable characterizing 
the sideward deflection of the nuclear matter emitted in non-zero
impact parameter collisions is the flow angle {\em i.e.} the angle between the 
direction of the collective sideward motion and the longitudinal beam axis.
Indeed in contrast to the flow parameter, 
${\Theta}_{\rm flow}$ is a measure of the overall emission direction of all
particles belonging to a given event and is not 
affected by thermal fluctuations. 
It follows therefore that the determination of this observable, because of 
its greater sensitivity, should help to better understand the origin of 
the flow and disentangle the different phenomena 
that may contribute to the observed effect.
The knowledge of ${\Theta}_{\rm flow}$ is, on the other hand, 
of great importance in the investigation of other phenomena such as
the out-of-plane squeeze-out effect~\cite{gut 90,cro 96,cro 97}
and two-particle interferometry~\cite{kot 95}.

Several methods were introduced to reconstruct 
this observable in high energy heavy-ion experiments. 
The sphericity method~\cite{gyu 82}, 
based on the diagonalization of the momentum flow tensor, allows 
the extraction of the flow angle as well as two aspects ratios characterizing 
the event shape (assumed to be ellipsoidal).
This procedure has the advantage that it provides an event-by-event shape
characterization. It deals, however, with the overall emission
pattern including the spectator component. Furthermore,
the sphericity analysis is strongly affected by the distorsions 
due to the effects of finite number of particles~\cite{dan 83}. 
An alternative shape analysis method, where the contribution of the
spectator matter can be removed, was proposed by Gosset et al~\cite{gos 90}.
The flow angle is adjusted, within this method, by fitting a simple 
anisotropic gaussian distribution to the triple differential momentum 
distributions of particles detected in the participant region. 
This procedure is nevertheless reliable only in the
case of low impact parameter collisions where the contribution of the 
spectator matter is not very important~\cite{bas 98}.

 We propose in the present article a new method which exploits 
 the saturation observed in the dependence of the flow parameter as a function
 of the fragment charge (Fig.~3 and 4).
 As discussed earlier 
 this saturation is due to the fact that IMFs are much more aligned along the 
 flow direction than do lighter particles : ``IMFs go with the 
 flow"~\cite{zha 95,gus 88}. 
 One expects therefore that if the phase space region occupied by IMFs is
 sufficiently elongated
 then their apparent flow angle $\Theta^{\rm app}_{\rm flow}$
 should give a good measure of the effective flow angle.

To investigate more quantitatively this idea, we have performed Monte-Carlo
simulations based on an anisotropically expanding thermal
source calculations, where particles with different masses share the same
thermal energy but their collective energy is proportional to their mass.
We assume that, at the freeze-out stage of the collision, the nuclear 
system is in local thermal equilibrium, {\em i.e.} same temperature
throughout the entire volume of the source. The latter is 
considered as a cylinder, in configuration space,
whose principal axis coincides with the flow axis.
The flow angle is introduced as a free parameter, 
$\Theta^{\rm input}_{\rm flow}$, in the simulations.
Particles are generated one by one independently of the influence on 
each other.
The momentum of a given fragment is considered as resulting from the 
superposition of a purely collective component (common velocity field
to all fragments) on the top of a thermal  
motion\footnote{details
about the simulations can be found in Ref.~\cite{cro 96}.} 

\begin{equation}
\vec{p}\;=\;\vec{p}^{\,\rm th}\;+\;\vec{p}^{\,\rm coll}
\end{equation}

The random thermal component $\vec{p}^{\,\rm th}$ is assumed to obey a 
Maxwellian distribution 

\begin{equation}
M(p^{\rm th},m)\;=\;{p^{\rm th}}^2
e^{- \sqrt{m^2+{p^{\rm th}}^2}/{\rm T}}
\end{equation}

where $m$ is the mass of the fragment and ${\rm T}$ is the temperature of the
source.

The collective component $\vec{p}^{\,\rm coll}$ is calculated assuming 
that at the moment of the explosion, 
the expansion velocity of a fragment increases linearly along 
the spatial coordinates towards the surface of the freeze-out volume 

\begin{equation}
v^{\rm coll}_{\rm i}\;=\;\frac{ r_{\rm i} }{ q_{\rm i} }\left. 
\sqrt{ \frac{ 2<{\rm E}^{\rm coll}_{\rm i}/{\rm A}> }{ 931.8 } }\; 
\right\}_{{\rm i}={\rm x,y,z}}
\end{equation}  

The fragment positions $r_{\rm i}$ (${\rm i}={\rm x,y,z}$), expressed in the source 
reference frame,
are randomly determined inside the cylindrical volume of the source  
assuming a uniform density at the freeze-out.
$<{\rm E}^{\rm coll}_{\rm i}/{\rm A}>$ is the mean collective 
energy per nucleon along the 
direction ${\rm i}$ and $q_{\rm i}$ is a normalization constant given 
by~\cite{ran 93} 

\begin{equation}
q_{\rm i}\;=\;\left. 
\sqrt{ \frac{ \sum_{{\rm j}=1}^{\rm N} r_{\rm i,j}^{2} }{\rm N} } \; 
\right\}_{{\rm i}={\rm x,y,z}}
\end{equation}

where ${\rm N}$ is the number of the particles generated in the simulations.

It is worth pointing out that the generation of full events is
beyond the scope of the present simulations. Our purpose here is to
simulate in a very simple way the phase space distributions of
fragments of different masses emitted in the mid-rapidity region.
The aim is to explore whether the apparent flow angle provided by IMF 
sideward flow data can be considered as an accurate measure of the
the effective flow angle.   

The simulations were carried out for semi-central events in the
MUL4 event bin at an incident energy of 
${\rm E} = 150~{\rm AMeV}$ where IMFs are 
copiously produced~\cite{ram 97,rei 97}.
We have used the following values as input parameters : 
${\rm T}=20~{\rm MeV}$, $<{\rm E}^{\rm coll}_{\rm x}/{\rm A}> 
= <{\rm E}^{\rm coll}_{\rm y}/{\rm A}> = 3.4~{\rm MeV}$,
$<{\rm E}^{\rm coll}_{\rm z}/{\rm A}> = 8~{\rm MeV}$ and 
$\Theta^{\rm input}_{\rm flow}={27.5}^{\circ}$.
Within this set of parameters, we could achieve a quite reasonable 
reproduction~\cite{cro 96} of the 
${\rm d}{\rm N}/{p_{\rm t}}{\rm d}{p_{\rm t}}$ distributions 
measured in the mid-rapidity region. 
The azimuthal anisotropy due to the squeeze-out effect~\cite{cro 96,bas 97}
was neglected ({\em i.e.} we assumed that 
$<{\rm E}^{\rm coll}_{\rm x}/{\rm A}> = <{\rm E}^{\rm coll}_{\rm y}/{\rm A}>$). 
The value of $\Theta^{\rm input}_{\rm flow}$ used in 
the calculations was taken equal to the apparent flow angle measured for 
${\rm Z}=7$ fragments.
The temperature parameter was found to affect only weakly the flow 
parameter of IMFs (less subject to thermal fluctuations). Its value
was constrained to match the experimental flow parameter of 
${\rm Z}=2$ particles.

In Fig.~5 we compare the experimental data (open circles), 
expressed in terms of the apparent flow angle versus the fragment charge, 
to the outcome of the simulations with (stars) and without (triangles) 
taking into account the detector filter~\cite{hol 00}. 
In the calculations, the mass of ${\rm A} \ge 4$ particles was
assumed to be equal to twice their charge ${\rm A}=2 \times {\rm Z}$. 
For ${\rm A} \le 3$ particles, the experimental isotopic ratios~\cite{pog 95}
were taken into account. As it can be seen, the filtered calculations
supply a fairly good quantitative reproduction of the observed charge 
dependence of the apparent flow angle. 
Note in particular that the saturation seen
in the data for the heaviest fragments is also present in the simulations. 
As the particles become heavier, the value of the apparent flow angle
converges toward 
the value of the effective flow angle $\Theta^{\rm input}_{\rm flow}$
(dotted line in Fig.~5) used as an input in the simulations. 
This strengthens the idea that a good estimate of the mean flow angle
can be extracted from the present measurements of IMF sideward flow.
It is also interesting to notice that the biases introduced by the FOPI 
apparatus do not affect significantly the apparent flow angle of IMFs.
At the incident energy considered here (${\rm E} = 150~{\rm AMeV}$), 
above ${\rm Z}=5$ (Fig.~5) filtered and 
unfiltered calculations give almost the same results.

\subsection{Scaling properties of the flow angle}

It is interesting to investigate now the dependences of the flow
angle on the impact parameter and the incident energy. To this end, we
applied the method described above to evaluate the flow angle from
the data. 
Practically, to quantify the constant limiting value from the 
${\rm Z-}$dependence of the apparent flow angle, 
we calculated the average over the $\Theta^{\rm app}_{\rm flow}$ 
values measured for the heaviest fragments in the saturation region.
The range of fragment charges included in the average, at each beam energy,
is indicated by the shaded area in Fig.~3. 
The results are displayed in Fig.~6 for the Au+Au reaction.
Error bars are the resulting uncertainties on the average value.

As is seen in the left-hand panel of Fig.~6 {\b{at an incident energy
of ${\rm E} = 400~{\rm AMeV}$}}, the flow angle is found 
to depend strongly on the collision centrality. 
We observe a monotonic increase of $\Theta_{\rm flow}$ with decreasing 
impact parameters. This trend is qualitatively in agreement with the 
predictions of several theoretical models~\cite{aic 87,sch 93,dan 95}.  
The latter models predict that $\Theta_{\rm flow}$ reaches 
${90}^{\circ}$ (with an oblate shape), 
in the limit of the most central collisions ($b \rightarrow 0$).
This very interesting issue cannot, however, be directly 
addressed on the basis of the present data as the procedure used to
evaluate $\Theta_{\rm flow}$ is not reliable for  
low impact parameter collisions. For relatively central events
($b_{geo} \le 3~{\rm fm}$), the dispersion on the reconstructed reaction 
planes is dramatically large and the effect on the apparent flow angle 
becomes difficult to be accounted for.
Investigations of c.m. polar angular 
distributions~\cite{rei 97,roy 97}   
are expected to be better suited for elucidating this still debated
question concerning the flow pattern in highly central collisions. 
           
The excitation function of the sideward flow was already investigated
in several experiments~\cite{cro 97,par 95,dos 86,cha 97,zha 90}.
However, in all these studies the sideward flow was expressed in terms
of the flow parameter or the directed transverse momentum
($p_x^{dir}$) 
which are, in contrast to $\Theta_{\rm flow}$, subject to the 
fluctuations introduced by the thermal motion. 
Here, we report for the first time on the incident energy
dependence of the flow angle. The latter observable  
being a measure of the ``pure" collective motion, its evolution with
the bombarding energy is expected to better reflect the collective response
of the nuclear system under different conditions.    
The experimental results are presented in the 
right-hand panel of Fig.6 for the Au+Au reaction over a very broad 
energy range going from $90$ to $800~{\rm AMeV}$. 
The data are shown for semi-central (MUL4 or PMUL4) events
where the directed sideward flow is close to its maximum.      

The observed trend (Fig.~6) is characterized by a steep rise of
$\Theta_{\rm flow}$ to a maximum at around ${\rm E} = 250-400~{\rm AMeV}$, 
followed by a moderate decrease as the bombarding energy increases further.
This is in complete disagreement with the scaling behaviour, 
{\em i.e.} constant flow angle, expected from 
ideal-fluid hydrodynamics~\cite{bon 87,sch 93}.
Quantitative calculations in the framework of one-fluid viscous 
hydrodynamics~\cite{sch 93} have shown that $\Theta_{\rm flow}$ scales
with the impact parameter almost 
independently of the incident energy (from $200$ to $800~{\rm AMeV})$. 
It was also found in these calculations
that the equation-of-state and the
viscosity do not influence significantly the value of
the flow angle. 
It seems therefore that, within this model, the flow angle is mainly 
governed by the collision geometry.

Deviations from scaling are visible in the data (Fig.~6)
at both low and high incident energies.
Such deviations are generally thought to signal~\cite{bon 87} the presence
of phenomena known to violate scaling such as the influence of the
equation-of-state and possible 
phase transitions~\cite{bon 87}. 
On the low energy end, the departure from scaling is 
particularly striking. One observes a drastic drop of the flow angle
below $150~{\rm AMeV}$.
This effect was already reported in an earlier publication 
where we have presented the excitation functions of the 
flow parameter~\cite{cro 96}. 
It is probably caused by a transition from predominantly attractive to 
repulsive forces resulting in a liquid-to-vapour phase transition.      
A rough linear extrapolation toward lower incident energies leads to an 
intersection energy where the sideflow vanishes, the so-called balance 
energy, of about $54~{\rm AMeV}$.
This value is somewhat lower than our previous evaluation in 
Ref.~\cite{cro 96,cro 97}. This is due to the fact that the data 
in Ref.~\cite{cro 96} were corrected for the resolution of the reaction 
plane using the method of Ref.~\cite{dan 85}. 
The latter method applied at low incident energies 
${\rm E} \le 120~{\rm AMeV}$, where the
fluctuations on the reaction plane are relatively large,
overestimates the correction factors and leads, therefore, to lower
flow values~\cite{pos 98}.

Changes in the nuclear viscosity, $\eta$, might also contribute to the 
sudden change of $\Theta_{\rm flow}$ at low energies. Below 
$\sim 100~{\rm AMeV}$, $\eta$ is relatively high because of the influence
of the Pauli principle. With increasing energies, $\eta$ is expected to drop
allowing therefore for a fast (sudden) buildup of pressure.  
This was quantitatively investigated in reference~\cite{dan 84}
where realistic viscosities, derived from Uehling-Uhlenbeck equations,
were used.

The origin of the smooth decline of $\Theta_{\rm flow}$ observed at high 
bombarding energies (Fig.~6) is not yet clearly understood. It might
be also attributed to a combination of several effects :
the anisotropy of the nucleon-nucleon interaction, 
the excitation of resonances and possible changes in the stiffness of the 
equation-of-state~\cite{ris 96}. 
Recent data from AGS~\cite{bar 97} and CERN/SPS~\cite{pos 97,app 98}
experiments at higher energies indicate relatively
small sideward flow effects. An estimate of 
$\Theta_{\rm flow} \sim 1/20 {\rm deg.}$ 
was reported from the Pb(158AGeV)+Pb data~\cite{pos 97}.    

Overall, the rise and fall of the sideward flow, although it is not yet
clearly understood,  reflects certainly important
changes in the properties of strongly interacting hot and dense 
hadronic matter.
At the low energy end, the onset of flow is indicative of a reduced 
pressure in the system and hence might testify to a 
liquid-gas phase transition. Similarly, the fall of flow on the
high energy side might also result from 
the possible transition towards the 
quark gluon plasma phase which is expected to soften 
the equation-of-state~\cite{ris 96}. 
The present data deserve, therefore, to be further investigated in 
conjunction with the higher energy data from AGS and 
SPS experiments. 
First attempts along this line have started very 
recently~\cite{rev 1}.  

\section{Summary and conclusion}

 In this work, a systematic study of the directed sideward flow 
 of light charged particles and intermediate mass fragments was 
 achieved with the FOPI detector at the SIS/ESR facility. 
 Three symmetric reactions, 
 Ru+Ru, Xe+CsI and Au+Au, were investigated at 
 beam energies between $90~{\rm AMeV}$ and $800~{\rm AMeV}$.
 The data were analysed according to the transverse momentum analysis
 method and corrections for the finite resolution of the reaction plane 
 reconstruction were applied. 

 The flow parameter exhibits the expected correlation with the collision
 centrality with a well defined maximum at an impact parameter around 
 ${\rm b} = 4.8~{\rm fm}$ and vanishing sideflow at both the low and 
 high ${\rm b}$ ends.  
 Consistent with earlier results~\cite{cha 97}, the sideward flow 
 scales according to the ${\rm A}_{\rm P}^{1/3}+{\rm A}_{\rm T}^{1/3}$ 
 scaling rule.

 The flow parameter is found to rise with
 the charge of the detected fragment up to ${\rm Z}=3-4$ and then turns into
 saturation for heavier fragments. This trend supports the concept
 of a collective motion and can be quantitatively accounted for by 
 simple simulations based on an anisotropically expanding thermal
 source. Based on these simulations, we have shown that the value of
 the flow parameter at saturation provides a good estimate of 
 the flow angle $\Theta_{\rm flow}$ in the participant region, 
 offering therefore a new method to evaluate this observable
 in the beam energy range considered here.  
 In contrast to the flow parameter, $\Theta_{\rm flow}$ has the advantage 
 to be unaffected by the influence of the thermal motion.  
 
 Applying this method, we have explored the dependences of the flow angle
 on both the collision centrality and the bombarding energy.
 In the impact parameter range
 ($b > 3~{\rm fm}$) where this method is applicable, 
 $\Theta_{\rm flow}$ rises monotonically as the collision becomes
 more central.
 This trend is qualitatively in agreement with the
 predictions of several theoretical models~\cite{aic 87,sch 93,dan 95}.
 The excitation function of $\Theta_{\rm flow}$ reveals
 clear deviations from the flat dependence expected from an 
 ideal hydrodynamical scaling.
 The data exhibit a steep rise of $\Theta_{\rm flow}$ to a maximum at
 around $250-400~{\rm AMeV}$, followed by a moderate decrease as the
 bombarding energy increases further.
 On the low energy end, the departure from scaling is
 particularly striking. The flow angle drops drastically 
 below $150~{\rm AMeV}$. This effect might be caused by a sudden
 decrease of the internal pressure due to the influence of the 
 attractive mean field. 
 A rough extrapolation toward lower incident 
 energies leads to an estimate of the balance energy 
 to about $54~{\rm AMeV}$.
 This value is consistent with the
 systematics of the balance energy~\cite{but 95}.  
 The origin of the smooth decrease of $\Theta_{\rm flow}$ observed 
 above $400~{\rm AMeV}$
 is not yet clearly understood. It might
 be attributed to a combination of several effects :
 the anisotropy of the nucleon-nucleon interaction, 
 the excitation of resonances and possible changes in the stiffness of the
 equation-of-state~\cite{ris 96}. 
 A quantitative evaluation of these effects would 
 require detailed comparisons with the predictions of 
 microscopic transport models. 
 Further studies in conjunction with the higher energy data from AGS and SPS 
 experiments should also shed some light on the mechanism responsible
 for the decline of flow at high incident energies. 

\section*{Acknowledgement}

This work was supported in part by the French-German agreement between
GSI and IN2P3/CEA and by the PROCOPE-Program of the DAAD.


\begin{table}
\caption{Measured cross section ($\sigma$), 
geometrical impact parameter
($b_{geo}$), reduced geometrical impact parameter
($b_{geo}/b_{max}$), accuracy on the determination of the reaction plane
($\sigma(\Delta\Phi_{\rm R})$) and the correction factors
($<cos(\Delta\Phi_{\rm R})>$)
for the three reactions under study.
The results are given for events in the MUL4 bin
at an incident energy of 
${\rm E}=400~{\rm AMeV}$.
The maximum impact parameter, $b_{max}$, was calculated assuming 
$r_0=1.20~{\rm fm}$.}
\vspace{0.5cm}
\begin{tabular}{cccc}
   & Ru+Ru & Xe+CsI & Au+Au \\ \hline
$\sigma\; ({\rm barn}) $  & 0.420  & 0.523 & 0.738 \\ \hline 
$b_{geo} \; ({\rm fm})$ & 2.83  & 3.12 & 3.61 \\  \hline
${b_{geo}}/{b_{max}}$ & 0.26 & 0.26 & 0.26 \\  \hline
$ \sigma(\Delta\Phi_{\rm R})\; (deg.)$ & 34.24  & 31.09 & 22.4 \\ \hline 
$ <cos(\Delta\Phi_{\rm R})>\;$ & 0.833  & 0.862 & 0.901 \\ 
\end{tabular}
\label{tab1}
\end{table}

\vspace{1.cm}

\begin{table}
\caption{Measured cross section ($\sigma$), mean geometrical impact parameter
($b_{geo}$ and $b_{geo}/b_{max}$), accuracy on the determination of the 
reaction plane ($\sigma(\Delta\Phi_{\rm R})$) and the correction factors
($<cos(\Delta\Phi_{\rm R})>$)
for Au+Au collisions at different beam energies.
The results are given for the MUL4 (PMUL4 for the 
$600$ and $800~{\rm AMeV}$ 
data) event class.
$b_{max}$ was calculated assuming $r_0=1.2~{\rm fm}$.}
\vspace{0.5cm}
\begin{tabular}{cccccccc}
${\rm E}~{\rm (AMeV)}$  & 90 & 120 & 150 & 250 & 400 & 600 & 800 \\ \hline
$\sigma~{\rm (barn)}$   & 0.577  & 0.720 & 0.608 & 0.753 & 0.738 & 
0.575 & 0.617 \\ \hline
$b_{geo}~{\rm (fm)}$ & 3.75  & 3.98 & 3.53 & 3.68 & 3.61 & 3.44 & 
3.58 \\ \hline 
$b_{geo}/b_{max}$ & 0.27 & 0.28 & 0.25 & 0.26 & 0.26 & 0.25 & 
0.23 \\ \hline
$ \sigma(\Delta\Phi_{\rm R})~{\rm (deg.)}$ & 46.2  & 39.03 & 33.7 & 24.7 & 
22.4 & 24.8 & 28.8 \\ \hline
$ <cos(\Delta\Phi_{\rm R})>\;$ 
& 0.552  & 0.758 & 0.838 & 0.893 & 0.901 & 0.885 & 0.847 \\
\end{tabular}
\label{tab2}
\end{table}

\vspace{1.cm}

\newpage



\begin{figure}
\caption{Normalized mean in-plane transverse momentum per 
nucleon (${<p_{\rm x}^{(0)}>}$) 
as a function of the normalized rapidity ($y^{(0)}$) for particles with 
charge ${\rm Z}= 1$ (full squares), 2 (open triangles), 3 (dots), 
4 (open squares), 5 (full triangles) and 6 (stars).
The data are shown for Au+Au (a), Xe+CsI (b) and Ru+Ru (c) reactions
for the same MUL4 centrality cut and at the same bombarding energy
of ${\rm E} = 400~{\rm AMeV}$. 
They are corrected for fluctuations of the reaction plane.
For the sake of clarity, 
an offset (${\rm const} = 0.7 \times ({\rm Z}-1)/10$) 
is applied to the data points 
corresponding to a given particle of charge ${\rm Z}$.
Error bars include statistical errors only.}
\label{fig1}
\end{figure}


\begin{figure}
\caption{Dependence of the normalized 
flow parameter on the geometrical impact parameter.
The data are shown at a bombarding energy of $400~{\rm AMeV}$.
Error bars are statistical errors multiplied by 
$\protect\sqrt{\chi ^2}$.
Geometrical impact parameters are deduced from the measured cross sections
assuming a sharp-cut-off approximation.}
\label{fig2}
\end{figure}


\begin{figure}
\caption{Normalized flow parameter
{\em versus} the charge of the detected fragment.
The data are shown for the Au+Au reaction at different beam energies,
under the MUL4 or PMUL4 (for $600$ and $800~{\rm AMeV}$) centrality cut.
Error bars are statistical errors multiplied 
by $\protect\sqrt{{\chi}^2}$. 
The solid lines are fits to the data using Fermi functions.   
The shaded area indicates the $Z$ range used to extract the average 
flow angle (see section IV.E) 
}
\label{fig3}
\end{figure}


\begin{figure}
\caption{Normalized flow parameter
scaled to $({\rm A}_{\rm P}^{1/3}+{\rm A}_{\rm T}^{1/3})$ {\em versus}
the charge of the detected fragment.
The results are shown for the three reactions under study at an incident
energy of $400~{\rm AMeV}$.
The MUL4 centrality cut was applied to all 3 reactions.
Error bars are statistical errors multiplied 
by $\protect\sqrt{{\chi}^2}$.}
\label{fig4}
\end{figure}


\begin{figure}
\caption{Apparent flow angle {\em versus} the fragment charge for
Au($150~{\rm AMeV}$)+Au semi-central (MUL4) collisions.
Error bars take into account statistical errors only.
The data (open circles) are compared to the results of anisotropic
expanding thermal source model simulations
with (stars) and without (triangles) the FOPI detector filter. 
The horizontal dotted line corresponds to the value of
$\Theta^{\rm input}_{\rm flow} = {27.5}^{\circ}$ used in the calculations.
See text for more details.}
\label{fig5}
\end{figure}


\begin{figure}
\caption{Left-hand panel : Mean flow angle as a function of the geometrical
impact parameter. Data are shown for the Au+Au reaction
at an incident energy of ${\rm E} = 400~{\rm AMeV}$.
Right-hand panel : Mean flow angle as a function of the bombarding energy.
Data are shown for Au+Au collisions under the MUL4/PMUL4 centrality cut
corresponding to mean geometrical impact parameters in the range
$3.5$ - $4~{\rm fm}$ (see Tab.2).
}
\label{fig6}
\end{figure}

\newpage

\begin{figure}[hhh]
\includegraphics{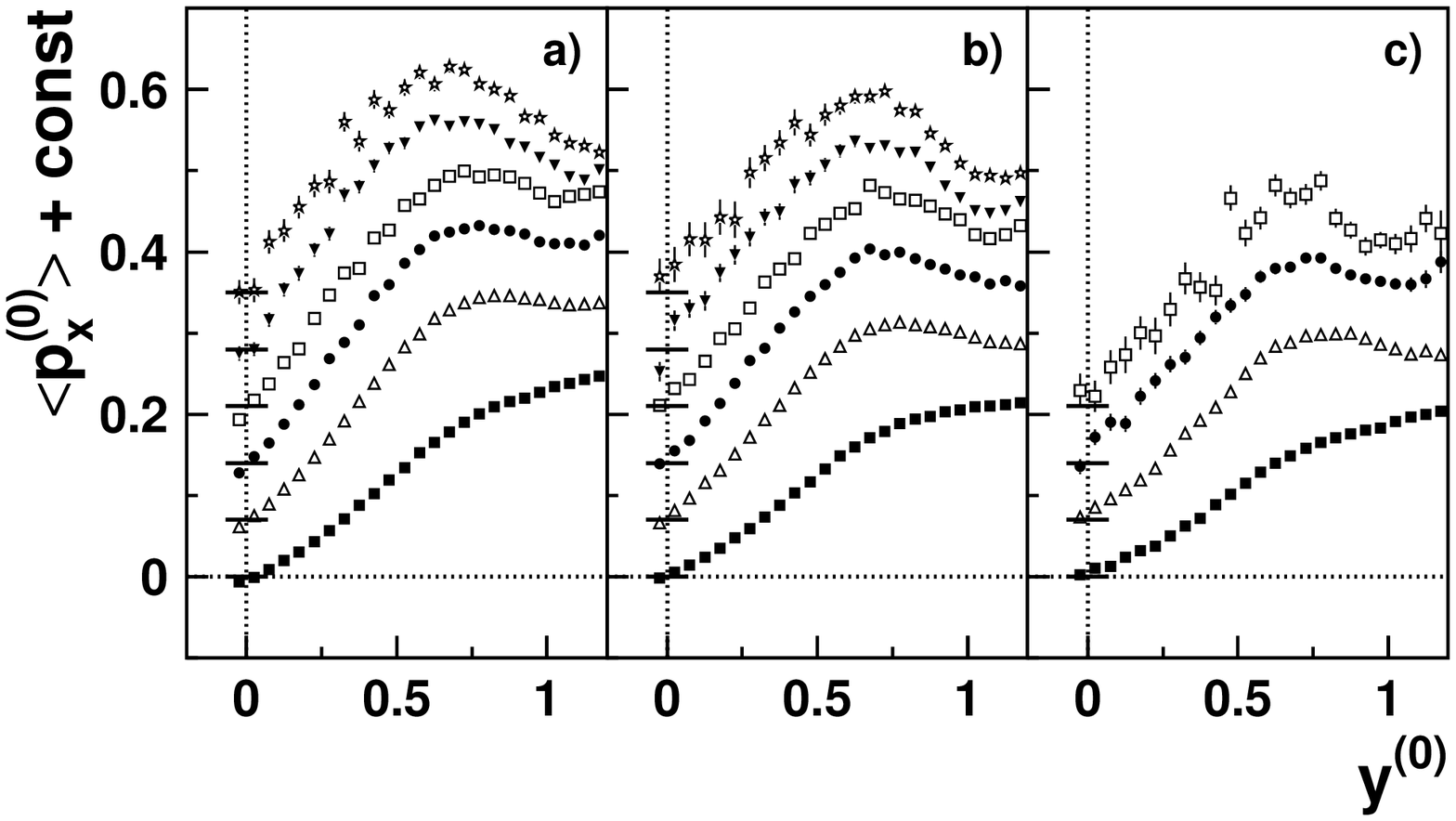}
\end{figure}
{\huge\bf\sf Figure 1}
\pagebreak

\begin{figure}[hhh]
\includegraphics{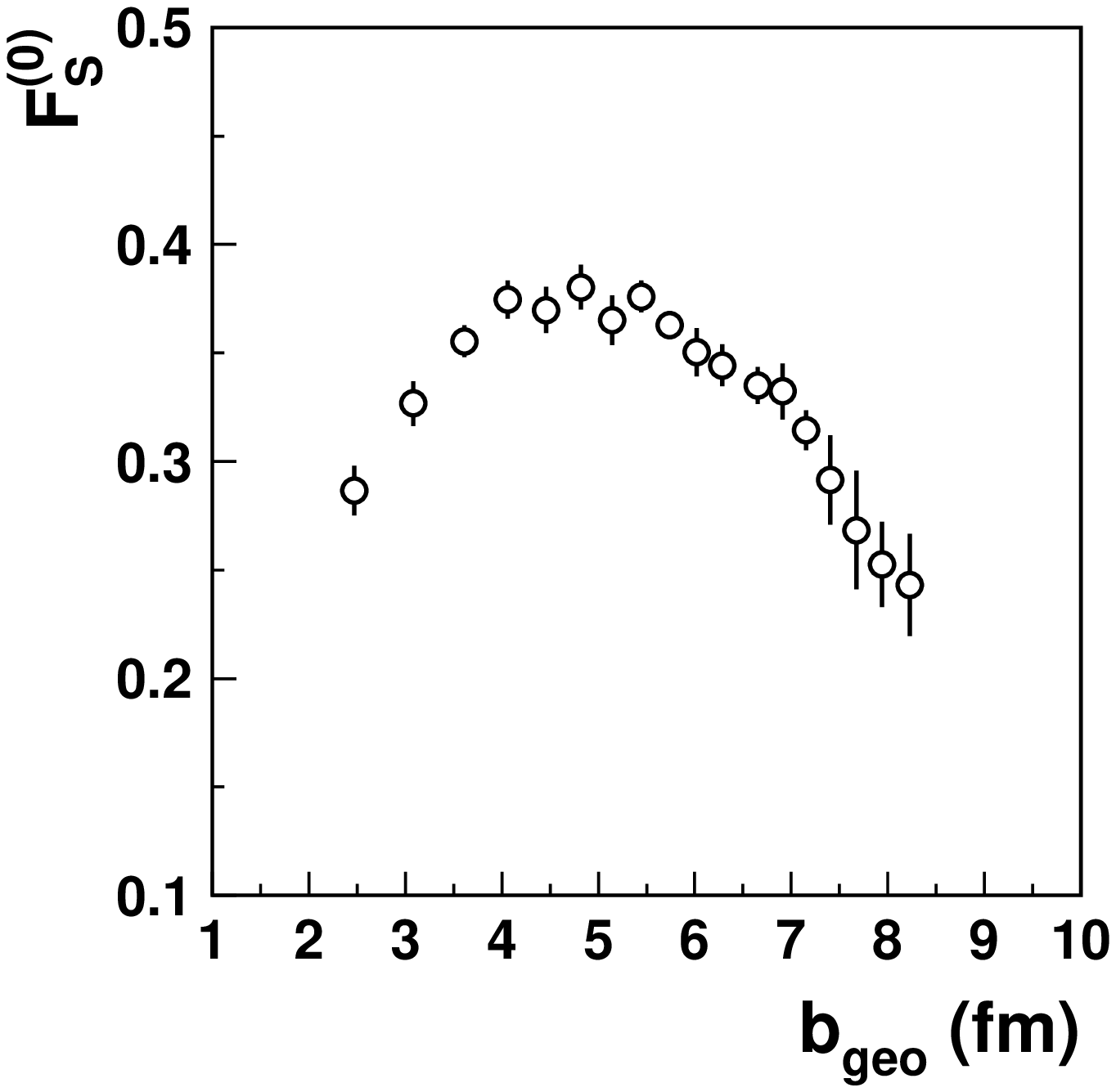}
\end{figure}
{\huge\bf\sf Figure 2}
\pagebreak

\begin{figure}[hhh]
\includegraphics{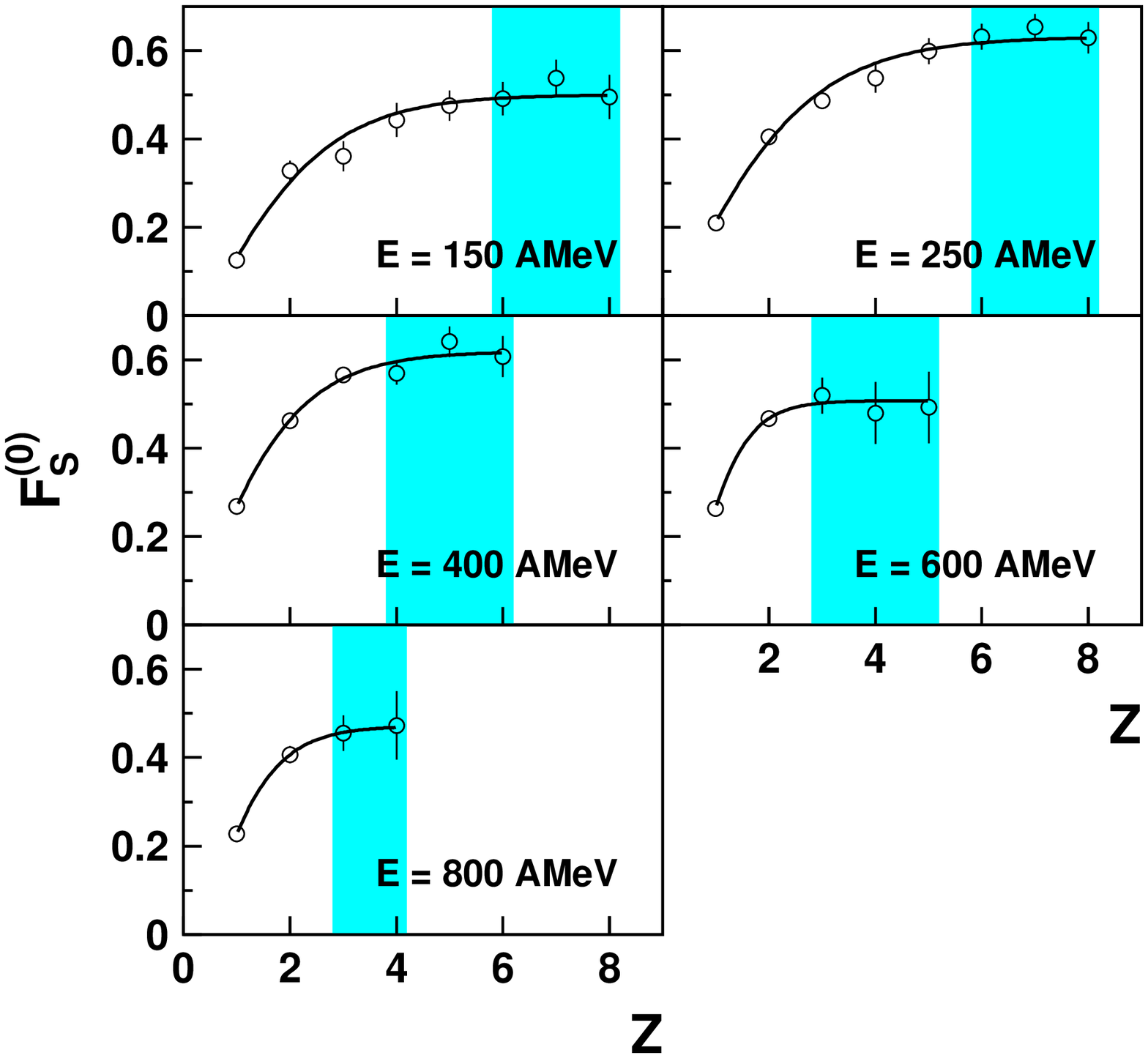}
\end{figure}
{\huge\bf\sf Figure 3}
\pagebreak

\begin{figure}[hhh]
\includegraphics{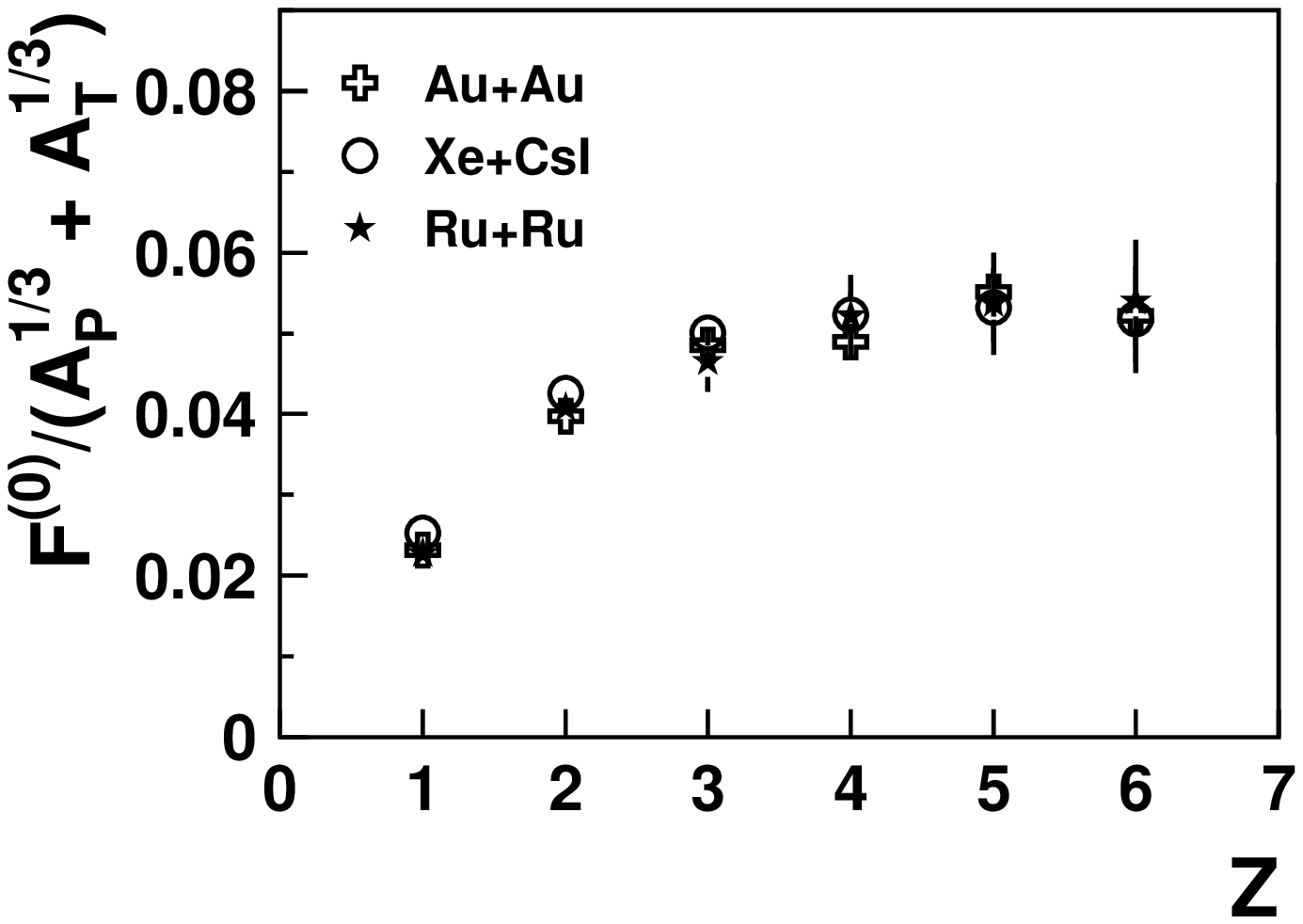}
\end{figure}
{\huge\bf\sf Figure 4}
\pagebreak

\begin{figure}[hhh]
\includegraphics{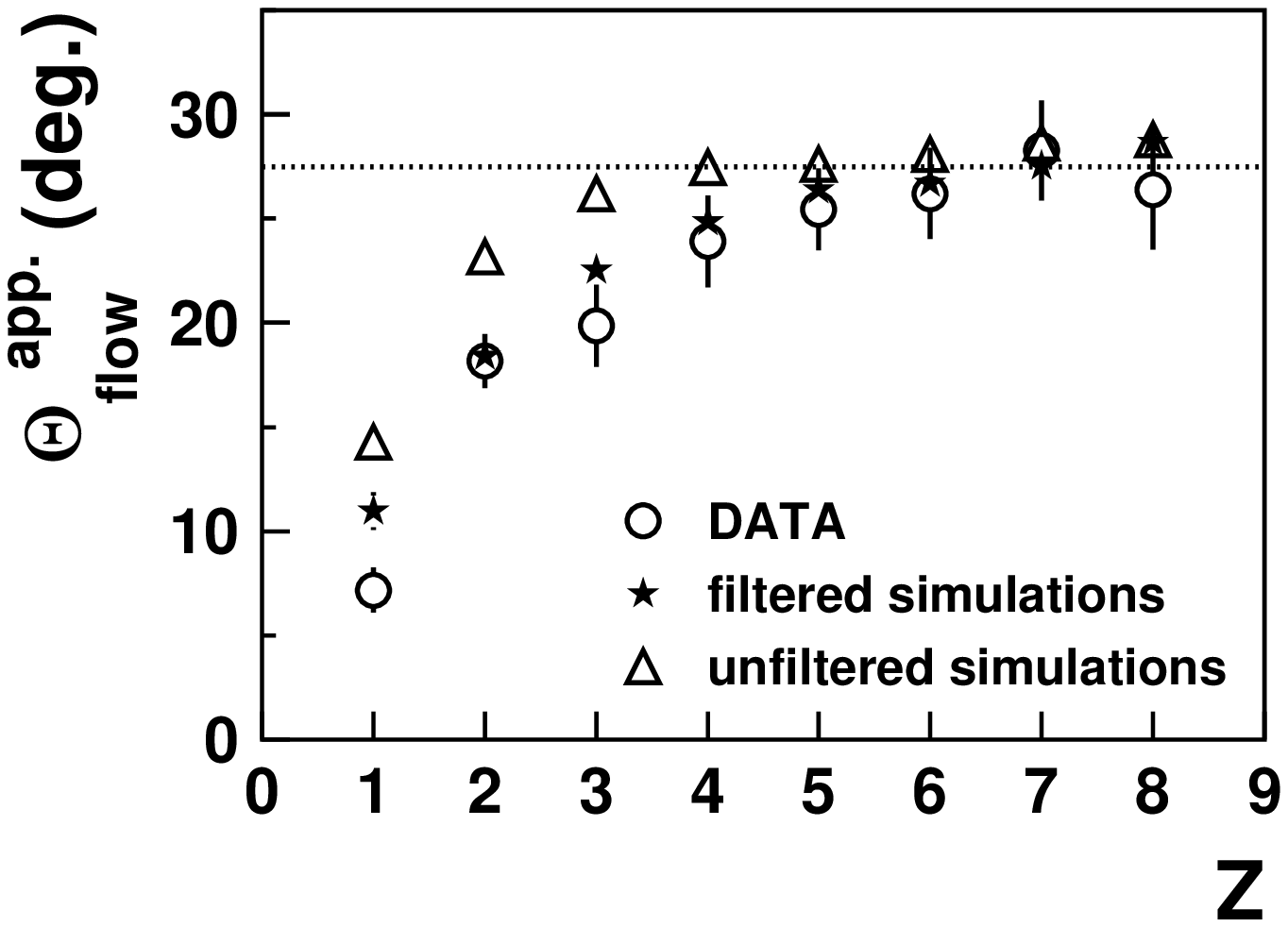}
\end{figure}
{\huge\bf\sf Figure 5}
\pagebreak

\begin{figure}[hhh]
\includegraphics{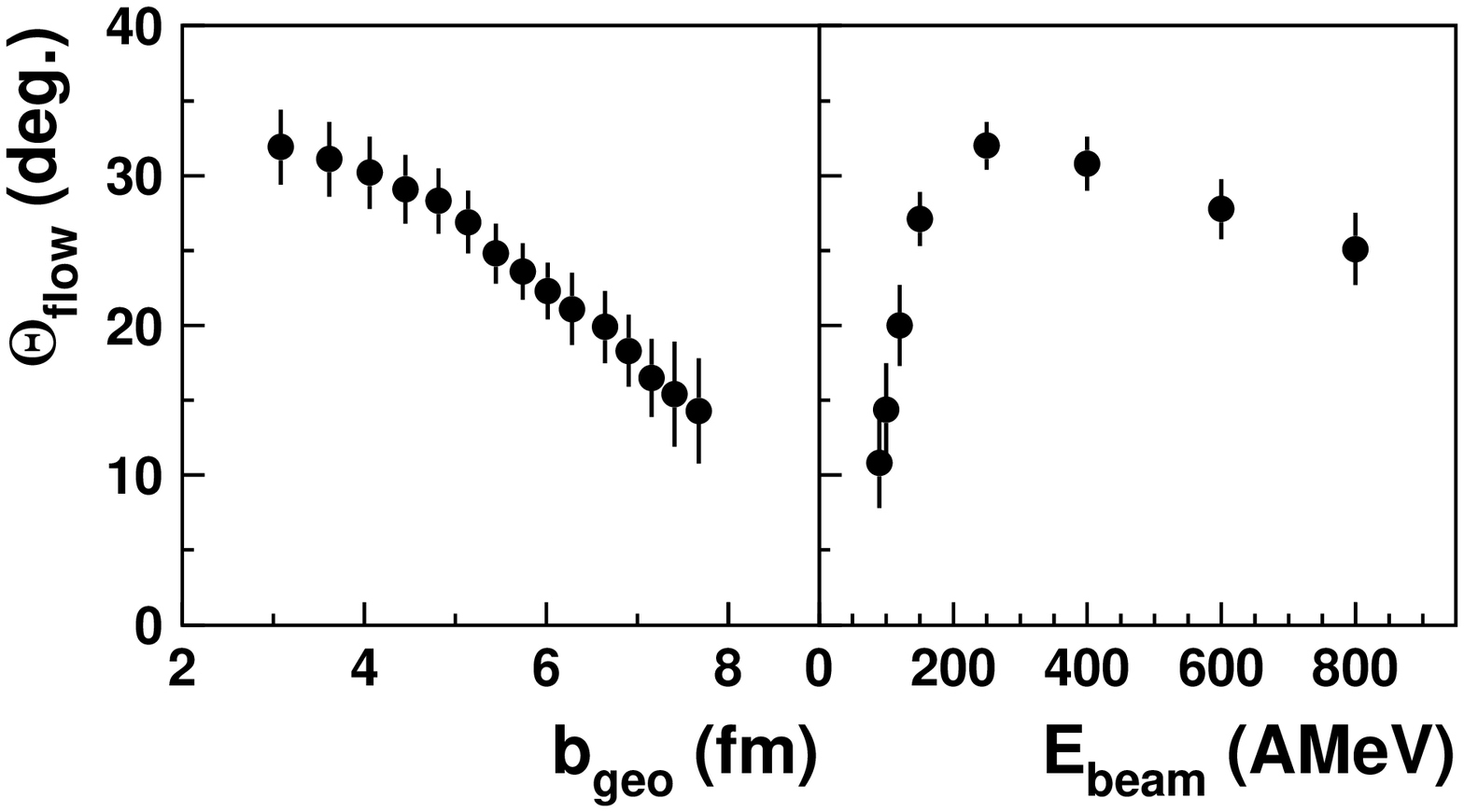}
\end{figure}
{\huge\bf\sf Figure 6}
\pagebreak

\end{document}